\documentclass[conference]{IEEEtran}
\usepackage{color}
\usepackage[ampersand]{easylist}
\usepackage{times}
\usepackage{booktabs}
\usepackage{comment}
\usepackage{listings}
\usepackage{mdframed}
 \usepackage{subcaption}
 \usepackage{graphicx}

\begin{document}

\definecolor{darkblue}{rgb}{0,0,.6}
\definecolor{darkred}{rgb}{.6,0,0}
\definecolor{darkgreen}{rgb}{0,.6,0}
\definecolor{red}{rgb}{.98,0,0}
\definecolor{gray}{rgb}{.6,.6,.6}

\title{Redesigning OP2 Compiler to Use HPX Runtime Asynchronous Techniques}
\author{Zahra Khatami$^{1,2}$, Hartmut Kaiser$^{1,2}$, and J. Ramanujam$^{1}$ \\$^{1}$Center for Computation and Technology, Louisiana State University\\$^{2}$The STE$||$AR Group, http://stellar-group.org
}

\maketitle

\begin{abstract}

Maximizing parallelism level in applications can be achieved by minimizing overheads due to load imbalances and waiting time due to memory latencies. Compiler optimization is one of the most effective solutions to tackle this problem. The compiler is able to detect the data dependencies in an application and is able to analyze the specific sections of code for parallelization potential. However, all of these techniques provided with a compiler are usually applied at compile time, so they rely on static analysis, which is insufficient for achieving maximum parallelism and producing desired application scalability. One solution to address this challenge is the use of runtime methods. This strategy can be implemented by delaying certain amount of code analysis to be done at runtime.

In this research, we improve the parallel application performance generated by the OP2 compiler by leveraging HPX, a C++ runtime system, to provide runtime optimizations. These optimizations include asynchronous tasking, loop interleaving, dynamic chunk sizing, and data prefetching. The results of the research were evaluated using an Airfoil application which showed a $40-50\%$ improvement in parallel performance.

\end{abstract}


\section{Introduction}

Unstructured grids are well studied and utilized in various application domains. OP2 provides a framework for the parallel execution of these unstructured grid applications on different multi-core/many-core hardware architectures~\cite{o1, o2}. The main goal of developing OP2 is to provide an abstraction level for users to parallelize their applications without having to worrying about architecture specific optimizations. This allows scientists to invest most of their time in understanding their domain problems, without learning details of new architectures, and still achieve efficient utilization of the available hardware. The framework is designed to achieve the near-optimal scaling on multi-core processors \cite{o4, o5}. However, as the compiler only has a static and defined access pattern \cite{comp1,comp2, comp3}, its analysis is not enough to obtain desired parallel scalability. In order to reach this goal, OP2 needs to be able to extract parallelism automatically at runtime. 

In this research, we propose different optimization methods that provide dynamic information for code generated by the OP2 compiler, including providing asynchronous task execution, interleaving different loops together, dynamically setting chunk sizes of different dependent loops based on each other, and prefetching data. These proposed techniques are implemented using HPX runtime system via redesigning the OP2 framework in a way that employs both  compiler's static analysis and dynamic runtime information. HPX is a parallel C++ runtime system that facilitates distributed operations and enables fine-grained task parallelism resulting in a better load balancing \cite{hpx1, hpx2}. It provides an efficient scalable parallelism by significantly reducing processor starvation and effective latencies while controlling overheads \cite{hpx3}.

A closer analysis of unstructured applications reveals that synchronization is only required between small tasks. Prevalent parallelization paradigms, however, coerce users to join all tasks together before proceeding to the next step in the application. In HPX, we can utilize the $future$ construct to allow every task to proceed as long as the values it depends on are ready \cite{future}.
This feature allows the HPX to relax the global barriers, enable flexibility, and improve the parallel performance of applications. In this research, HPX uses $futures$ based techniques to develop a new task execution strategy for codes generated by the OP2 compiler which is the basis for asynchronous tasking and interleaving loops.

In order to control the overheads introduced by the creation of each task, it is important to control the amount of work performed by each task. This amount of work is known as the \textit{chunk size} \cite{future, hpx4}. In addition, to properly interleave loops it is important for each loop to have very similar execution times which allows the waiting time between the execution of each loop to be minimal. We propose to address these two obstacles by creating a new execution policy which will dynamically control the chunk sizes during the application's execution. In addition, we also propose to create a new cache prefetcher that aids in prefetching data for each time step to reduce memory accesses latencies. This method is implemented in such a way that data of the next iteration step is prefetched into the cache memory using a prefetching iterator called in each iteration within a loop. The main difference between this method and the other existing methods is that HPX implementation combines a thread based prefetching method with the asynchronous task execution, which results in having asynchronous execution while prefetching data of all the containers within a loop. 


To our knowledge, we present a first attempt of redesigning OP2 to utilize the runtime techniques for improving performance of the parallel unstructured grid applications. The combination of these proposed techniques should yield a more portable and performant software stack for unstructured grid applications and enable the applications to properly scale to a higher level of parallelism compared to the existing OP2 implementation. The results evaluated in Section \ref{res} show that the parallelization performances are improved by around $40-50\%$ for an Airfoil application. The remainder of this paper is structured as follows: Section~\ref{op2} briefly introduces OP2; Section \ref{hpx} introduces a \textit{dataflow} object in HPX; Section \ref{imp} shows the details of the \textit{dataflow} implementation with the new execution policy within OP2; Section \ref{pref} presents the prefetching method implemented in one of the HPX parallel algorithms, and Section~\ref{res} evaluates the the scaling speedup of the experimental tests. The conclusions and the future works can be found in Section~\ref{con}.

\section{OP2}
\label{op2}

OP2 is an active library that provides a parallel execution framework for unstructured grid applications on different multi-core/many-core hardware architectures~\cite{o1}. It utilizes a source-to-source translator for generating code which targets different hardware configurations \cite{o2,o4,o3}. The code can be transformed easily into different configurations such as serial, multi-threaded using OpenMP and CUDA, or heterogeneous which utilizes MPI, OpenMP, and CUDA \cite{o4}. In this section, we first walk through a simple OP2 code to show its implementation details and then we introduce the Airfoil application which is used as a case study for this research. 

\subsection{Simple Code Implementation with OP2}

\begin{figure} [h!]
\begin{center}
\centering
\includegraphics[width=0.6\columnwidth]{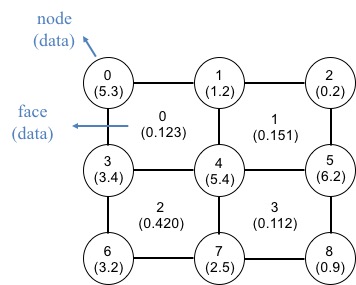}
\caption {The mesh represented data layouts provided with OP2.}
\label{OP2b}
\end{center}
\end{figure}

This section generally shows how unstructured grids are defined with OP2. The OP2 API handles the data dependencies by providing mesh represented data layouts. The provided framework is defined based on sets, data on sets, mapping connectivity between the sets, and the computation on each set \cite{o2,o7}. Sets can be nodes, edges or faces. In these unstructured grids, the connectivity information is used to specify different mesh topologies. Figure~\ref{OP2b} shows a mesh example that includes nodes and faces as sets. The value of data associated with each set is shown below each set and the mesh is represented by the connections between them. 

OP2 API for the mesh in figure \ref{OP2b} is shown as follows, which is the C/C++ API and defines 12 edges and 9 nodes:

    \begin{lstlisting}[basicstyle=\footnotesize]
op_set nodes;
op_decl_set(9, nodes, "nodes");
op_set edges;
op_decl_set(12, edges, "edges");
    \end{lstlisting}

The mapping that declares the connection between $2$ nodes is defined as follow:

    \begin{lstlisting}[basicstyle=\footnotesize]
int edge_map[28]={0,1,1,2,2,5,5,4,4,3,3,6,6,7,
		  7,8,0,3,1,4,2,5,3,6,4,7,5,8}

op_map pedge;
op_decl_map(edges,nodes,2,edge_map,pedge,"pedge")
    \end{lstlisting} 

\textit{op\_decl\_map} shows that each edge is mapped on two different nodes. The values of each node and face are assigned as follow:

    \begin{lstlisting}[basicstyle=\footnotesize]
float valueFace[4]={0.123, 0.151, 0.420, 0.112};
float valueNode[9]={5.3,1.2,0.2,3.4,5.4,6.2,3.2,2.5,0.9};
op_dat data_face;
op_decl_dat(face,1,"float",valueFace, 
	    data_face, "data_face");
op_dat data_node;
op_decl_dat(node, 1, "float", valueNode, 
	    data_node, "data_node");
    \end{lstlisting} 
    
These sets and meshes are used to define a loop over a given set. The more details about OP2 design and performance analysis can be found in \cite{o1} and \cite{o7}, which shows that all 
unstructured grid applications can be easily described with sets and meshes as shown in the above example. These methods place no restriction on the algorithm and they allow the programmer to choose unique operations on each set.

\subsection{Airfoil Application}

In this research, we study an Airfoil application, which is a standard unstructured mesh finite volume computational fluid dynamics (CFD) code, presented in \cite{o8}, for the turbomachinery simulation and consists of over $720K$ nodes and about $1.5$ million edges. As described in \cite{o8} and \cite{hpx7}, it consists of five parallel loops: \textit{op\_par\_loop\_save\_soln}, \textit{op\_par\_loop\_adt\_calc}, \textit{op\_par\_loop\_res\_calc}, \textit{op\_par\_loop\_bres\_calc}, \textit{op\_par\_loop\_update}, shown in figure \ref{l1}. All of the computations on each set are implemented within these loops by performing operations of the user's kernels defined in a header file for each loop: \textit{save\_soln.h}, \textit{adt\_calc.h}, \textit{res\_calc.h}, \textit{bres\_calc.h} and \textit{update.h}. Each argument passed to each loop is generated based on data values used with \textit{op\_arg\_dat}.

\begin{figure}
\begin{mdframed}
    \begin{lstlisting}[basicstyle=\scriptsize]
op_par_loop_save_soln("save_soln", cells,
  op_arg_dat(data_a0,...),...,
  op_arg_dat(data_an,...);
  
op_par_loop_adt_calc("adt_calc",cells,
  op_arg_dat(data_b0,...),...,
  op_arg_dat(data_bn,...);
  
op_par_loop_res_calc("res_calc",edges,
  op_arg_dat(data_c0,...),...,
  op_arg_dat(data_cn,...);

op_par_loop_bres_calc("bres_calc",bedges,
  op_arg_dat(data_d0,...),...,
  op_arg_dat(data_dn,...);
  
op_par_loop_update("update",cells,
  op_arg_dat(data_e0,...),...,
  op_arg_dat(data_en,...);
  
    \end{lstlisting}
    \end{mdframed}
    \caption{\small{Five loops used in \textit{Airfoil.cpp} for saving old data values, applying computation, and updating each data value.}}
    \label{l1}
\end{figure}

Figure \ref{fig:o1} demonstrates \textit{op\_par\_loop\_save\_soln} that applies \textit{save\_soln} on cells based on the arguments generated with \textit{op\_arg\_dat} using \textit{p\_q} and \textit{p\_qold} data values. The function \textit{op\_arg\_dat} creates an OP2 argument based on the information passed to it. These arguments explicitly indicate that how each of the underlying data can be accessed inside a loop: \textit{OP\_READ} (read only), \textit{OP\_WRITE} (write) or \textit{OP\_INC} (increment to avoid race conditions due to indirect data access) \cite{o1}. More details can be found in \cite{o2} and \cite{o3}. 

\begin{figure}
\begin{mdframed}
    \begin{lstlisting}[basicstyle=\footnotesize]
op_par_loop_save_soln("save_soln", cells,
op_arg_dat(p_q,-1,OP_ID,4,"double",OP_READ),
op_arg_dat(p_qold,-1,OP_ID,4,"double",OP_WRITE));
    \end{lstlisting} 
\end{mdframed}
\caption{\small{\textit{op\_par\_loop\_save\_soln} represents one of the loops used in an Airfoil application.}}
    \label{fig:o1}
\end{figure}

\begin{figure}
\begin{mdframed}
    \begin{lstlisting}[basicstyle=\footnotesize]
    
void op_par_loop_save_soln(char const *name, 
  op_set set, op_arg arg0, op_arg arg1)
{
.
.
.
#pragma omp parallel for
for(int blockIdx=0; blockIdx<nblocks; blockIdx++) 
{
  int blockId  = //based on the blockIdx
  int nelem    = //based on the blockId 
  int offset_b = //based on the blockId
        
  for (int n=offset_b; n<offset_b+nelem; n++) 
  {
   .
   .
   .
   save_soln(...); //user's kernel
  }}
 } 
    \end{lstlisting}
    \end{mdframed}
    \caption{\small{\textit{\#pragma omp parallel for} is used for a loop parallelization for an Airfoil application.}}
    \label{l2}
\end{figure}

The loop parsed with OP2 in figure \ref{l2} illustrates how each cell updates its data value by accessing \textit{blockId}, \textit{offset\_b}, and \textit{nelem} data elements. The arguments are passed to the \textit{save\_soln} user kernel subroutine, which does the computation for each iteration of an inner loop from \textit{offset\_b} to \textit{offset\_b+nelem} of each iteration of an outer loop from \textit{0} to \textit{nblocks}. Also it illustrates that OpenMP is used for the parallel processing within a node. It is important to note that the outputs of the computations shown in figure \ref{l1} cannot be passed to the outside of the loop, therefore, the current OP2 design doesn't provide a method for interleaving loops together. This creates implicit global barrier after each loop as the threads inside the loop must wait to synchronize before exiting the loop \cite{mpi3}. Barriers, naturally, impede optimal parallelization by causing the parallel threads and processes to wait. In order to solve this problem, this research sets out to optimize the performance of code generated by the OP2 compiler using the HPX runtime. The source-to-source code translator of OP2 is written in Matlab and Python \cite{o3}. In this research, its Python source-to-source code translator is modified to automatically generate the parallel loops using HPX library calls. 

\section{HPX}
\label{hpx}

In this research different dynamic optimizations are proposed for improving the performance of code generated by the OP2 compiler that are implemented using HPX runtime system, which has been developed to overcome limitations such as global barriers and poor latency hiding \cite{hpx2, hpx3} by embracing new ways of coordinating parallel execution, controlling synchronization, and implementing latency hiding utilizing Local Control Objects (LCO) \cite{hpx7, hpx8}. These objects have the ability to create, resume, or suspend a thread when triggered by one or more events.
LCOs provide traditional concurrency control mechanisms such as various types of mutexes, semaphores, spinlocks, condition variables and barriers in HPX. These objects improve the efficiency of an application by permitting highly dynamic flow control as they organize the execution flow, omit global barriers, and enable thread execution to proceed as far as possible without waiting. More details about LCO design and its performance can be found in \cite{hpx3, hpx6, hpx11}.

The two implementations of LCOs most relevant to this research are the $future$ construct and the \textit{dataflow} template. HPX provides a multi-threaded, message-driven, split-phase transaction, and distributed shared memory programming model using $futures$ and \textit{dataflow} based synchronization on the large distributed system architectures, which are explained in the following sections. 

\subsection{Future}
$future$ is a computational result that is initially unknown but becomes available at a later time~\cite{future}. The goal of using $future$ is to let every computation proceed as far as possible. Using $future$ enables  threads to continue their executions without waiting for the results of the previous steps to be completed, which eliminates the implicit global barrier at the end of the execution of an OpenMP parallel loop. $future$ based parallelization provides the rich semantics for exploiting higher level parallelism available within each application that may significantly improve its scalability. 

\begin{figure} 
\begin{center}
\centering
\includegraphics[width=0.5\columnwidth]{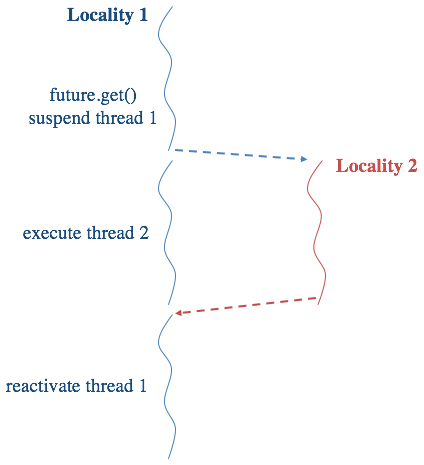}
\caption {\small{The principle of the operation of the $future$ in HPX. Thread $1$ is suspended only if the results from locality $2$ are not readily available. Thread $1$ accesses the $future$ value by performing \textit{$future.get()$}. If results are available, Tread $1$ continues to complete the execution.}}
\label{f4}
\end{center}
\end{figure}

Figure~\ref{f4} shows the scheme of the $future$ performance with 2 \textit{localities}, where a \textit{locality} is a collection of processing units (PUs) that have access to the same main memory. It illustrates that the other threads do not stop their progress even if the thread, which waits for the value to be computed, is suspended. Threads access a $future$ value by performing \textit{$future.get()$}. When the result becomes available, the $future$ resumes all HPX suspended threads waiting for that value. It can be seen that this process eliminates the global barrier synchronizations, as only those threads that depend on the $future$ value are suspended. With this scheme, HPX allows asynchronous execution of the threads.

\subsection{Dataflow Object}

\begin{figure} 
\begin{center}
\centering
\includegraphics[width=0.5\columnwidth]{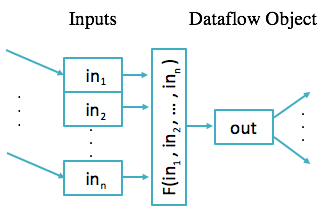}
\caption {A \textit{dataflow} object encapsulates a function $F(in_1,in_2,...,in_n)$ with $n$ inputs from different data resources. As soon as the last input argument has been received, the function $F$ is scheduled for an execution.}
\label{o3}
\end{center}
\end{figure}

\textit{dataflow} object provides a powerful mechanism for managing data dependencies without the use of global barriers \cite{hpx9, hpx1}. Figure \ref{o3} shows the schematic of a \textit{dataflow} object, which encapsulates a function $F(in_1,in_2,...,in_n)$ with $n$ future or non-future inputs from different data resources. If an input is a future, then the invocation of the function will be delayed. Non-future inputs are passed through. A \textit{dataflow} object waits for a set of $futures$ to become ready and as soon as the last input argument has been received, the function $F$ is scheduled for the execution \cite{hpx6}. Because the \textit{dataflow} object returns a $future$, its result can be fed to other objects in the system including other \textit{dataflows}. These chained $futures$, by their nature, represent a dependency tree that automatically generates an execution graph. This graph is executed by the runtime system as each nodes dependencies are meet. 
As a result, \textit{dataflow} minimizes the total synchronization by scheduling new tasks as soon as they can be run instead of waiting for entire blocks of tasks to finish computation.


\section{Implementing Dataflow in OP2}
\label{imp} 

\begin{figure} 
\begin{mdframed}
    \begin{lstlisting}[basicstyle=\footnotesize]    
using hpx::lcos::local::dataflow;
using hpx::util::unwrapped;  
  
//automatically returns the argument as a future
return dataflow(unwrapped([&](data_a,...){
    //same as original op_arg_dat
    return arg; 
    }
  }),data_a,...);  
    \end{lstlisting}
     \end{mdframed}
    \caption{\small{\textit{op\_arg\_dat} is modified to create an argument as a $future$ that is passed to a function through \textit{op\_par\_loop} shown in figure \ref{l1}.}}
    \label{o2}
           
\end{figure}

In this section, the new method is proposed for parallelizing loops generated with OP2, which is based on \textit{dataflow} implementation that solves the current challenges of OP2. In this method, the OP2 API is modified in such a way that \textit{op\_arg\_dat} used in each loop in figure \ref{l1} produces an argument as a $future$ for \textit{dataflow} object inputs. Figure \ref{o2} shows the modified \textit{op\_arg\_dat}, in which \textit{data\_a,...} expressed at the last line of the code invokes a function only once all of them get ready. \textit{unwrapped} is a helper function in HPX, which unwraps the $futures$ and passes along the actual results. This implementation also generates an output argument as a $future$ and as a result, all of the arguments of each loop in figure \ref{l1} are passed as a $future$ to the kernel function through \textit{op\_par\_loop}. 

\subsection{Parallelizing Loops Using \textit{for\_each}}
\label{parloop}
Parallelizing loops and controlling chunk sizes are implemented by using \textit{for\_each} algorithm and \textit{persistent\_auto\_chunk\_size} as an \textit{execution\_policy} respectively. In figure \ref{l6b}, \textit{dataflow} is implemented with \textit{for\_each} for the loop in figure \ref{l2}, that aids to parallelize the outer loop. \textit{for\_each} is one of the HPX parallel algorithms that is able to automatically control the chunk size during the execution by determining number of the iterations to be run on each HPX thread. Moreover, HPX is able to execute loops in sequential or in parallel by applying \textit{execution\_policies}, which are briefly described in Table \ref{tab:policies} \cite{hpx6}. The concept of the \textit{execution\_policy} developed in HPX is used to specify the execution restrictions of the work items, in which calling with a sequential execution policy makes the algorithm to be run sequentially and calling with a parallel execution policy allows the algorithm to be run in parallel \cite{hpx8}. 

\begin{table}
\centering
\begin{tabular}{lll}
       \toprule
       Policy & Description & Implemented by \\
       \hline
       \textit{seq} & sequential execution & Parallelism TS, HPX \\
       \textit{par} & parallel execution & Parallelism TS, HPX \\
       \textit{par\_vec} & parallel and   & Parallelism TS \\
       & vectorized execution & \\
       \textit{seq(task)} & sequential and  & HPX \\
       & asynchronous execution & \\
       \textit{par(task)} & parallel and & HPX \\
       & asynchronous  execution & \\
       \bottomrule
       
    \end{tabular}
    \caption{\small{The execution policies implemented in HPX.}}
    \label{tab:policies}
\end{table}

\begin{figure}
\begin{mdframed}
    \begin{lstlisting} [basicstyle=\footnotesize]  
    
hpx::shared_future<op_dat> op_par_loop_save_soln(char const * name, op_set set,
    hpx::future<op_arg> arg0,
    hpx::future<op_arg> arg1)
{
using hpx::lcos::local::dataflow;
using hpx::util::unwrapped;    
//automatically returns output as a future
return dataflow(unwrapped([&save_soln]
(op_arg arg0, op_arg arg1){
   .
   .
   .
  auto r=boost::irange(0, nblocks);
  hpx::parallel::for_each(policy, 
  r.begin(), r.end(),
  [&](std::size_t blockIdx){  
        
  for (int n=offset_b; n<offset_b+nelem; n++)
  {
   .
   .
   .
   save_soln(...);
  }
  return arg1;
}),arg0,arg1);
}
    \end{lstlisting}
    \end{mdframed}
    \caption{\small{Implementing \textit{for\_each} within \textit{dataflow} for the loop parallelization in OP2 for the loop in figure \ref{l2}. It makes the invocation of a loop asynchronous by returning output as a  $future$.  \textit{dataflow} allows automatically creating execution graph, which represents a dependency tree.}}
    \label{l6b}      
\end{figure}

\begin{figure} 
\begin{mdframed}
    \begin{lstlisting}[basicstyle=\footnotesize]     
    
p_qold = 
	op_par_loop_save_soln("save_soln",cells,
	op_arg_dat(p_q,-1,OP_ID, 
			4,"double",OP_READ),
	op_arg_dat(p_qold,-1,OP_ID,
			4,"double",OP_WRITE));
    \end{lstlisting}
    \end{mdframed}
    \caption{\small{\textit{Airfoil.cpp} is changed while using \textit{dataflow} for the loop parallelization in OP2. \textit{p\_qold} is returned as a $future$ from each kernel function after calling \textit{op\_par\_loop}.}}
    \label{l3c}
    \end{figure}

\begin{figure}
\begin{mdframed}
\centering
    \includegraphics[width=\columnwidth]{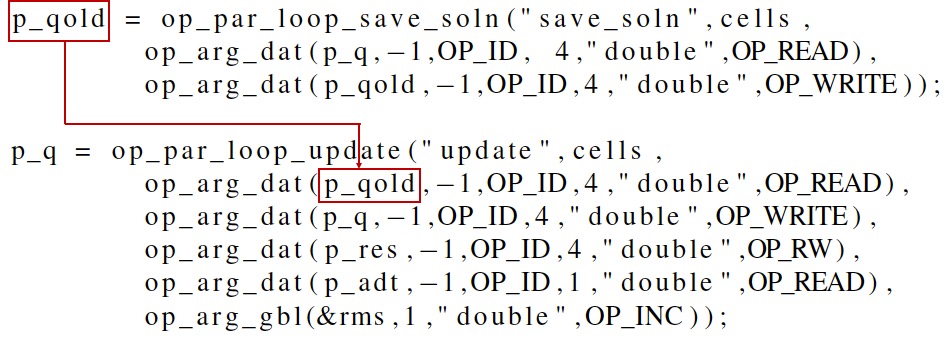}
    \end{mdframed}
    \caption{\small{The proposed method makes OP2 able to interleave thses two loops together by passing \textit{p\_qold} output of \textit{op\_par\_loop\_save\_soln} as an input argument for \textit{op\_par\_loop\_update.}}}
    \label{inter1}
\end{figure}    

\begin{figure} 
\begin{center}
\centering
\includegraphics[width=\columnwidth]{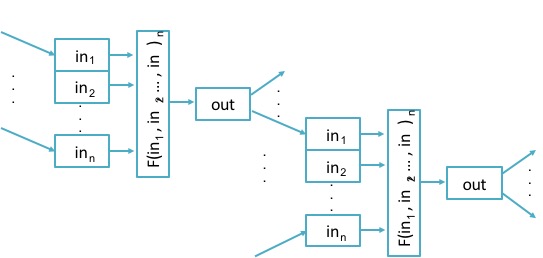}
\caption {\textit{dataflow} provides a way for interleaving execution of different loops together by generating output as a $future$ and passing all inputs as $future$s as well.}
\label{data3}
\end{center}
\end{figure}

Figure \ref{l6b} also illustrates that $arg0$ and $arg1$, which are created as a $future$ with \textit{op\_arg\_dat} using \textit{p\_q} and \textit{p\_qold} respectively, are passed as a $future$ within a loop. This loop will be executed only if these arguments get ready. Then, the output argument, which is $arg1$ in this example, is passed as a $future$ to the outside of the loop and it is stored within \textit{p\_qold} shown in figure \ref{l3c}. This method is implemented to all of the loops in figure \ref{l1}, and as a result, each kernel function returns an output argument as a $future$. The loop execution may depend on the results of the other previous loops. So by this method, the results of the loops can be passed as $future$ inputs to the other loops, which makes OP2 able to interleave different loops. For example, \textit{p\_qold} value updated in \textit{op\_par\_loop\_save\_soln} is used as an input argument for \textit{op\_par\_loop\_update} as shown in figure \ref{inter1}, which using this proposed method makes it able to interleave this two loops together by passing output of \textit{op\_par\_loop\_save\_soln} as an input argument for \textit{op\_par\_loop\_update}. 

Figure \ref{data3} shows generally that by implementing proposed method, the $future$ output of each loop passed as an input of the other loops makes OP2 able to interleave different loops together at runtime.  As a result, if the loops are not dependent on each other, they can be executed without waiting for the previous loops to complete their tasks, however, if they depend on the parameters from the previous loops, they will wait until the previous loops complete their processes. This proposed method removes the unnecessary barrier synchronizations between different loops and execute them asynchronously. 

\subsection{Controlling Chunk Sizes}
\label{chunk}

\begin{figure}
\centering
\begin{subfigure}{0.49\columnwidth}
\includegraphics[width=\columnwidth]{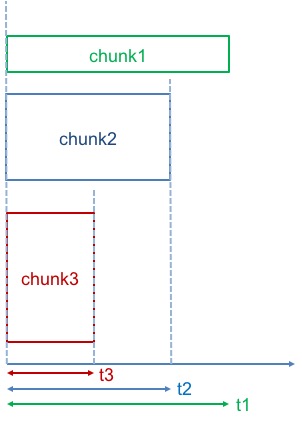}
\caption {Chunk sizes with different execution time}
\label{chunk1}
\end{subfigure}
\begin{subfigure}{0.49\columnwidth}
\includegraphics[width=\columnwidth]{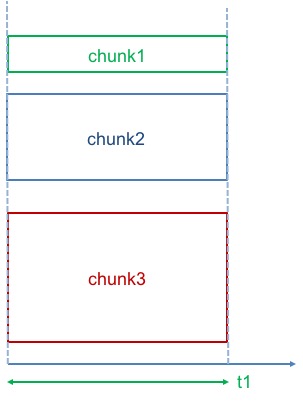}
\caption {Chunk sizes with the same execution time}
\label{chunk2}
\end{subfigure}

\caption{\small{Setting chunk sizes of different dependent loops based on each other.}}
\label{fig:test}
\end{figure}

As it is explained in section \ref{parloop}, figure \ref{data3} shows how \textit{dataflow} provides a way of interleaving execution of different loops together. In a case of having dependent loops, the execution of each chunk in a loop depends on the execution of the chunks in the previous loop. By using \textit{par} as an execution policy, different chunks with different execution time regardless of the chunk sizes of the other loops are determined for each loop shown in figure \ref{chunk1}, which may increase the waiting time between them. So for decreasing this waiting time, the execution time of each chunk in these dependent loops should be the same. For this purpose, the new execution policy is proposed in this section, named \textit{persistent\_auto\_chunk\_size}, that makes all chunk sizes of different loops having same execution time as shown in figure \ref{chunk2}. In this policy, the chunk size of the first loop is determined automatically with \textit{for\_each} algorithm. Then the chunk sizes of each second and third loops are determined based on the execution time of the chunk in the first loop. As a result, all chunks of all these three loops will have the same execution time. It should be note that \textit{chunk1}, \textit{chunk2} and \textit{chunk3} have different sizes but with the same execution time. 

\section{HPX Data Preftecher}
\label{pref}

Data prefetching is one of the methods for reducing memory accesses latencies by calling data required for the next step into the cache \cite{pr2}. The simplest form of the cache prefetching can be implemented by prefetching cache line of the next iteration as soon as the current cache line is referenced \cite{pr1, pr3}. Hardware, software and thread prefetching are different traditional techniques for this purpose.

Various hardware prefetching methods has been proposed that one of them is using one-block-lookahead (OBL) scheme \cite{pr6}. In this method, the blocks $i + 1$, $i+2$, ..., and $i+n$ are prefetched whenever the block $i$ is brought to the cache that results in reducing cache misses significantly. Creating reference prediction table \cite{pr7, pr8} is another method to limit unnecessary prefetching and to predict the future memory references. However, one of the big challenges exists in most of these hardware prefetching methods is that the prefetcher uses the past access pattern by considering data stream, which cannot handle an irregular access pattern.

In the software prefetcher method, the prefetching data is implemented by using prefetch directives in the code. One of the problem of this method is that these prefetching instructions are inserted with programmer or compiler into the applications, which has the high probability of the cache miss occurrences. Another problem is introducing additional overhead for executing these prefetch instructions. There has been many developments proposed for optimizing this technique that mostly are obtained by prefetching pointer-based data structures \cite{pr6, pr5}. Mowry's algorithm \cite{sf1} is one of the recent prefetching optimization that defines the affine array-references as the prefetching candidates within an inner-most loop, performs the loop unrolling, and creates the multiple memory references within a loop. As a result, the exact missing instance is prefetched, which avoids the unnecessary prefetching and reduces prefetching overheads. Jump pointer prefetching \cite{pr7, pr8} is another proposed software prefetching approach, which is implemented by inserting additional pointers into a dynamic data structure for connecting non-consecutive elements within a loop. This technique allows prefetching data by creating pointer chain and results in overlapping fetching process of multiple elements simultaneously. However, this technique also has the difficulty in handling sequences of the irregular data accesses \cite{pr1}.

Thread based prefetching method is usually preferred over the software / hardware prefetching methods, since it precomputes the load addresses accurately and it is able to follow more complex patterns compared to the other methods \cite{pr3}. This technique executes an application in the prefetcher thread context and brings data of the next cache line into the shared cache before the main thread accesses it. However, the scaling can be degraded with this method because of 

\begin{enumerate}
\item cache misses: the prefetcher could make slower progress than the main thread, and
\item global barriers: a global barrier is needed to synchronize the prefetcher with the main thread \cite{pr2,pr3,pr6}. 
\end{enumerate}

\begin{figure}
\centering
    \includegraphics[width=0.7\columnwidth]{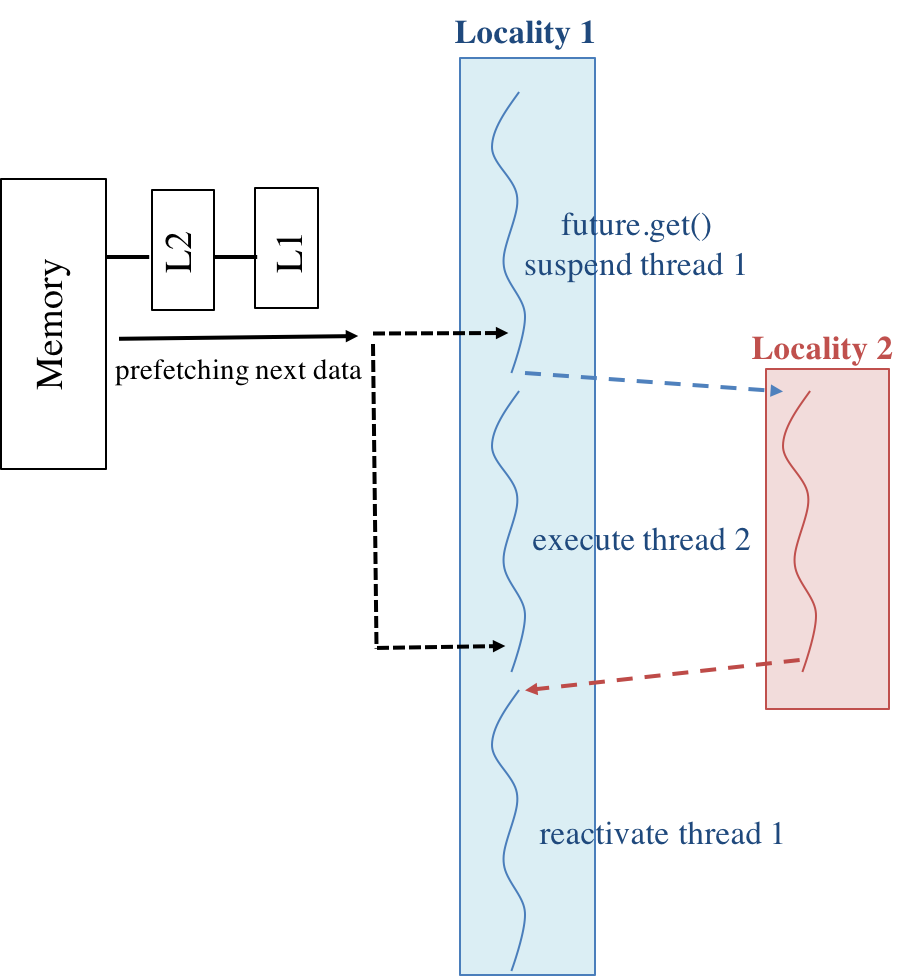}
    \par\caption{\small{Data of the next iteration step is prefetched into the cache memory with the prefetching iterator called in each iteration within the \textit{for\_each}}}
    \label{l5}
\end{figure}

\begin{figure} 
\begin{mdframed}
    \begin{lstlisting} [basicstyle=\footnotesize]    
auto ctx=hpx::parallel::make_prefetcher_context(
	loop_range.begin(), loop_range.end(),
	prefetch_distance_factor,
	container_1,container_2, ...,container_n);

hpx::parallel::for_each(policy,
        ctx.begin(), ctx.end(),
        [&](std::size_t i)
        {
           container_1[i] = ...;
           container_2[i] = ...;
            .
            .
            .
           container_n[i] = ...;
         });      
    \end{lstlisting}
      \end{mdframed}
    \caption{\small{The prefetching method used in \textit{for\_each}. The prefetching iterator in \textit{for\_each} is called by using \textit{ctx\_begin}, which is the struct that references to all container in the loop.}}
    \label{l111}
\end{figure}

In this section, the new prefetching method is introduced in HPX that combines a thread based prefetching with an asynchronous task execution. The main goal of this method is not only to reduce the memory accesses latencies, but also to relax the global barriers, which results in a better parallel performance. 

Figure \ref{l5} shows the scheme of using $future$ and the proposed prefetching iterator, which makes HPX to have the asynchronous execution while prefetching data of all the containers within a loop of the next step in to the cache memory in each iteration. Moreover, HPX is able to prefetch data in sequential or in parallel with applying \textit{execution\_policy} described in Table \ref{tab:policies}. This method is added to the method explained in section \ref{parloop} to decrease the memory access latencies while parallelizing  loops.

Figure \ref{l111} shows the details of the prefetching method implementation within \textit{for\_each}. The program execution is divided into several chunks within \textit{for\_each} and its iterator is developed to prefetch data of the next chunk size in either sequential or in parallel. The prefetching iterator is initialized with calling constructer of \textit{make\_prefetcher\_context} and it is executed by using \textit{ctx.begin()}, which is the struct that references to all containers used in a loop and \textit{loop\_range} is the range, in which the loop is executed. One of the feature of this prefetcher is that it works with any data types even in a case of having different type for each container. The distance between each two prefetching operations is computed based on the value of \textit{prefetch\_distance\_factor}. In order to increase the effectiveness of the prefetcher and to decrease the relative cost, \textit{prefetch\_distance\_factor} is designed to be determined based on the length of the cache line. As a result, within each prefetcher distance, data of all containers of the next time step are prefetched in each iteration by calling this prefetching iterator. The experimental results of optimizing OP2 performance with HPX discussed in this paper are presented in the next section.

\section{Experimental Results}
\label{res}

In this section, we evaluate the experimental results of our work by comparing our proposed framework to OP2's current design. The main goal of this section is to illustrate that dynamic information obtained at runtime and static information obtained at compile time are both necessary to provide sufficient optimizations for optimal performance. The proposed methods studied in the previous sections  are evaluated here. The experiments are executed on the test machine with two Intel Xeon E5-2630 processors, each with 8 cores clocked at 2.4GHZ and 65GB. Hyper-threading is enabled. The OS used by the shared memory system is 32 bit Linux Mint 17.2. and HPX 0.9.99 is used here.

\subsection{Asynchronous Task Execution Provided with Dataflow}

\begin{figure} 
\begin{center}
\centering
\includegraphics[width=\columnwidth]{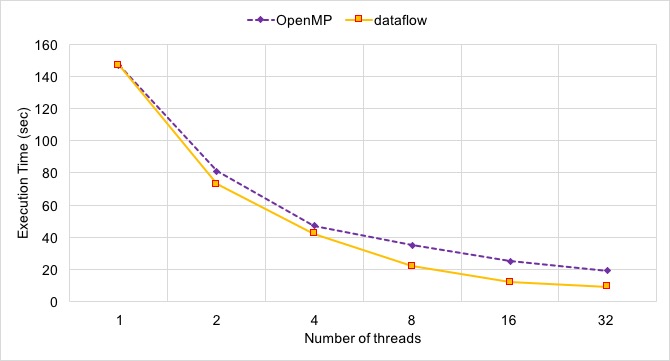}
\caption {\small{Comparison results of the execution time between \textit{dataflow} and \textit{\#pragma omp parallel for} used for an Airfoil application.}}
\label{f6}
\end{center}
\end{figure}

\begin{figure} 
\begin{center}
\centering
\includegraphics[width=\columnwidth]{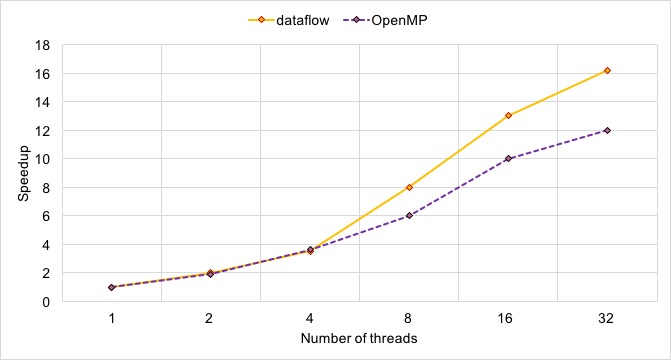}
\caption {\small{Comparison results of the strong scaling between \textit{dataflow} and \textit{\#pragma omp parallel for} used for an Airfoil application. This comparison result illustrates a better performance for \textit{dataflow} for a larger number of threads, which is due to the asynchronous task execution. \textit{dataflow} automatically generates an execution tree, which represents a dependency graph and allows an asynchronous execution of the functions. Hyperthreading is enabled after 16 threads.}}
\label{f2}
\end{center}
\end{figure}

Figure \ref{f6} shows the execution time of an Airfoil application using \textit{\#pragma omp parallel for} and \textit{dataflow}, which illustrates that HPX and OpenMP has approximately the same performance on $1$ thread. We are however able to improve parallel performance in using \textit{dataflow} for more number of threads. For the speedup analysis, we use strong scaling, for which the problem size is kept the same as the number of cores are increased. Figure \ref{f2} shows the strong scaling comparison results that illustrates a $33\%$ better performance for \textit{dataflow} due to the asynchronous task execution, the use of $futures$, and interleaving different dependent loops together. As described in section \ref{hpx}, \textit{dataflow} automatically generated an (implicit) execution tree, which represents a dependency graph that results in removing unnecessary global barriers and improving scalability of the parallel applications.

\subsubsection{Controlling Chunk Sizes}

\begin{figure} 
\begin{center}
\centering
\includegraphics[width=\columnwidth]{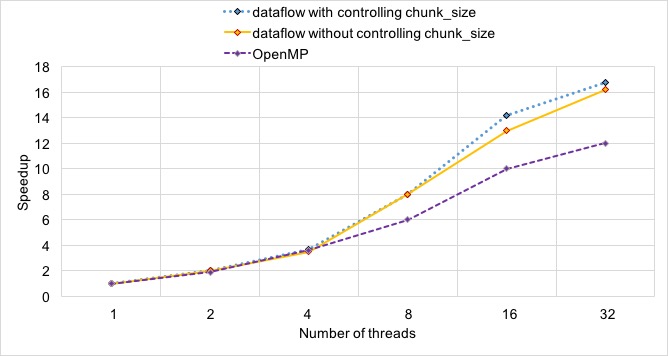}
\caption {\small{Comparison results of strong scaling using \textit{dataflow} with/without setting chunk sizes of different dependent loops based on each other. Hyperthreading is enabled after 16 threads.}}
\label{res3}
\end{center}
\end{figure}

\begin{figure} 
\begin{center}
\centering
\includegraphics[width=\columnwidth]{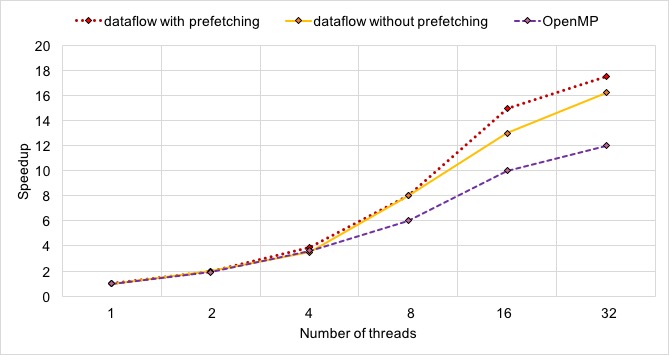}
\caption {\small{Comparison results of a \textit{dataflow} performance by using proposed prefetching method. It shows that the speedup is increased by around $45\%$ with prefetching data within a loop. Hyperthreading is enabled after 16 threads.}}
\label{dataflow2}
\end{center}
\end{figure}

\begin{figure} 
\begin{center}
\centering
\includegraphics[width=\columnwidth]{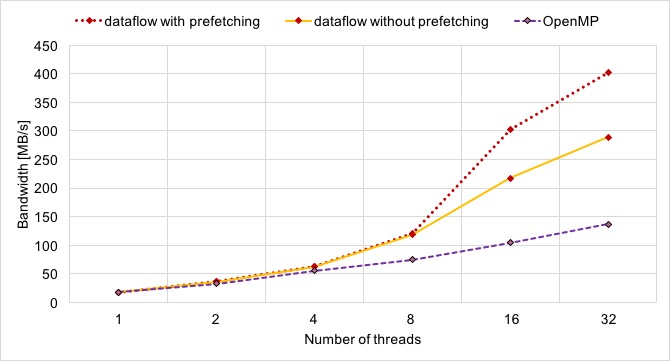}
\caption {\small{The data transfer rate of implementing hpx::for$\_$each using standard random access iterator versus prefetching iterator within a \textit{dataflow}. Hyperthreading is enabled after 16 threads.}}
\label{band1}
\end{center}
\end{figure}

\begin{figure} 
\begin{center}
\centering
\includegraphics[width=\columnwidth]{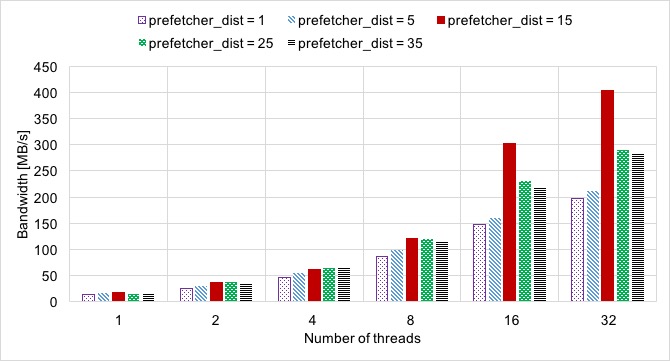}
\caption {\small{The data transfer rate of using prefetching iterator for different prefetching distances. Hyperthreading is enabled after 16 threads.}}
\label{band2}
\end{center}
\end{figure}

In this section, the chunk sizes of different loops are set by considering chunk sizes determined in the previous loops. Since \textit{dataflow} enables the compiler to interleave different loops together, the execution of each chunk in each loop depends on the execution of the chunks in the previous loops. So using \textit{persistent\_auto\_chunk\_size} makes the execution time of each chunks in these loops to be the same, which decreases the waiting time between them. Figure \ref{res3} shows the improvement in the performance of \textit{dataflow} method by using \textit{persistent\_auto\_chunk\_size} as an execution policy within the loops. For an instance, with $32$ threads, the improvement is obtained by about $40\%$.

For further parallelization performance improvement, data prefetching proposed in section \ref{pref} is implemented in the \textit{dataflow} method and its results are evaluated in the next section.

\subsection{Prefetching Data}

The proposed prefetching method is applied on the \textit{dataflow} method and its performance is shown in figure \ref{dataflow2}. This method takes advantage of the asynchronous execution while prefetching  data within a loop of the next step in to the cache memory in each iteration step. These results illustrate that the parallel performance of \textit{for\_each} is improved by an average of $45\%$, which confirms the successful process of avoiding cache misses with implementing HPX prefetcher iterator. The bandwidth rate comparison of these results are also shown in figure \ref{band1}.

The results of the parallel performance of the prefetching iterator measurements with different \textit{prefetch\_distance\_factor} are shown in figure \ref{band2}. It can be seen that for the very large distances, data prefetching cannot improve the parallel performance. On the other hand, very small prefetcher distances causes more data to be prefetched, which becomes more expensive. This cost dominates the gains from prefetching and impedes scaling. It is illustrated that $prefetch\_distance\_factor~=~15$ for an Airfoil application improves the parallel performance significantly. These results show the good scalability achieved by HPX and indicates that it has the potential to continue to scale on a larger number of threads.

\section{Conclusion}
\label{con}

In this research, we present an implementation of the OP2 compiler that employs HPX runtime techniques to efficiently and automatically parallelize unstructured grid applications to achieve desired parallel scalability. The results illustrate that using both dynamic information provided at runtime and the static information provided at compile time are necessary to obtain a higher parallelism level in the applications. 

In the proposed framework, OP2 is able to automatically produce  data dependencies based on arguments that are passed into the loops at compile time and, by using HPX parallelism methods, the generated loops can be executed asynchronously. In this framework, we propose different optimization methods that make OP2 execute tasks asynchronously, interleave different loops together, efficiently control the chunk sizes of different dependent loops based on each other, and prefetch data into the cache before its actual access. These proposed methods improved the overall performance of an Airfoil application by $40-50\%$.

In future research, we plan to improve runtime optimizations with information from the compiler. Since runtime information is often speculative, solely relying on it doesn't guarantee maximizing parallelization performance. In general, the parallelization performance of an application depends on the values measured at runtime and the related transformations such as loop skewing and loop scheduling performed at compile time. Collecting the outcome of the static analysis performed by the compiler could significantly improve the runtime performance. \\~\\

\noindent
\textbf{Acknowledgements}

We would like to thank Adrian Serio from Center for Computation and Technology at Louisiana State University for the invaluable and helpful comments and suggestions to improve the quality of the paper. This works was supported by NSF awards 1447831.

\bibliography{Diss_bib}
\bibliographystyle{unsrt}

\end{document}